# Universal precursor seismicity pattern before locked-segment rupture

Hongran Chen[a, b], Siqing Qin[a, b, c], Lei Xue[a, b], Baicun Yang[a, b, c] and Ke Zhang[a, b, c]

[a]Key Laboratory of Shale Gas and Geoengineering, Institute of Geology and Geophysics, Chinese Academy of Sciences, Beijing, China. [b]Innovation Academy for Earth Science, Chinese Academy of Sciences, Beijing, China. [c]College of Earth and Planetary Sciences, University of Chinese Academy of Sciences, Beijing, China.

*Siqing Qin

**Email:** qsqhope@mail.iggcas.ac.cn

**Author Contributions**

S. Q. initiated the study. S. Q. and H. C. conceived the idea of this study. L. X., B. Y. and K. Z. analyzed the data. H. C. and L. X. interpreted the results. H. C. wrote the original manuscript. All authors discussed the results and revised the manuscript.

**This PDF file includes:**

    Main Text

    Figures 1 to 6

    Tables 1 to 2

**Abstract**

Despite the enormous efforts towards searching for precursors, no precursors have exhibited real predictive power with respect to an earthquake thus far. Seismogenic locked segments that can accumulate adequate strain energy to cause major earthquakes are very heterogeneous and less brittle; progressive failures of the locked segments with these properties can produce an interesting seismic phenomenon: a characteristic earthquake and a sequence of smaller subsequent earthquakes (pre-shocks) always arise prior to another characteristic earthquake within a well-defined seismic zone and its current seismic period. Applying a mechanical model and magnitude constraint conditions, we show that two adjacent characteristic earthquakes reliably occur at the volume-expansion and peak-stress points of a locked segment. Such a seismicity pattern has occurred in 62 seismic zones worldwide, suggesting that the pattern applies universally. Both the precursor pattern and the model quantifying it permit the prediction of certain characteristic earthquakes in a seismic zone.

**Significance**

Reliable earthquake prediction, whether short- or long-term, is vital for reducing earthquake hazards. However, owing to poor understanding regarding earthquake mechanisms, reliable predictions could not be made until now. Herein, we show that (i) locked segments in seismic zones are seismogenic structures that dominate the generation of tectonic earthquakes; (ii) a characteristic earthquake at the volume-expansion point of a locked segment represents a discernible precursor for the next characteristic earthquake at the peak-stress point of the segment; (iii) the earthquake generation mechanism can arise from brittle failures of stressed locked segments, which macroscopically rupture one-by-one in the order of bearing capacity from low to high; and (iv) the evolution of characteristic earthquakes follows a deterministic rule, thereby ensuring their predictability.

**Main Text**

**Introduction**

Reliable earthquake prediction, although very important for preventing and mitigating earthquake disasters, has remained elusive. The key to satisfactory prediction is the identification of reliable earthquake precursors. In recent decades, several studies have been reported on the seismological, geodetic, geophysical, geochemical, and hydrological precursors to earthquakes (1); however, most of these are physically irrelevant or statistically insignificant. Thus far, no single precursor signal has exhibited real predictive power concerning a major earthquake (1, 2). We have analyzed the seismic-generation process to identify a reliable signal.

Tectonic earthquakes are widely understood as the brittle failure of local rock accompanied by a rapid release of energy owing to fault and plate movement (3, 4). A major earthquake (e.g., a $M_W 8.0$ quake) is usually preceded by an intermittent long-term process involving multiscale cracking events in a specified region, whereas the acoustic emissions (AEs) generated from a stressed rock specimen are associated with small-scale cracking events within a short period. Nevertheless, given the similarities in the sequences and statistical distributions of earthquakes and AE events (4-7), a primary goal of laboratory studies about AE monitoring is to identify a precursor signal that could be used for earthquake prediction.

Numerous laboratory-scale rock-mechanics experiments (8-11) have shown that a macroscopic rupture at the peak-stress point (PSP) of a rock specimen subjected to compression or shear is preceded by the emergence of a volume-expansion point (VEP), as shown in Fig. 1*A*. The VEP defines the boundary between the stable and unstable crack growth phases (8). This pattern is essential to the evolution of damage in heterogeneous rocks and provides a physical basis for predicting major earthquakes (12). Lockner and Byerlee (13) noted that the spatial and temporal clustering or unstable propagation of micro-cracks beginning at the VEP is capable of causing a supra-exponential rise in AE activity that coincides with anomalies in seismic indicators, such as wave velocity (14) and electrical and magnetic signals (3). However, these seeming precursors to macroscopic rupture (Fig.1*B*) have not been consistently observed in laboratory tests as the mechanical behavior of a stressed rock specimen and AE activity pattern heavily depend on its heterogeneity, size, shape, and the applied loading conditions. Furthermore, the expansion-related precursor phenomena observed in laboratory settings are not useful for realistic earthquake prediction because recorded spatial and temporal clusters of earthquakes typically do not correspond clearly with a subsequent large event. We argue that when searching for a reliable seismic precursor, one must first understand the physical construct that accumulates high strain energy, and only then can one formulate the related earthquake mechanism.

Several geoscientists (7, 15, 16) have recognized that heterogeneous seismogenic faults comprise both fault gouges and strong local segments that resist slip. Each of the strong segments with high bearing capacity (high strength and/or large scale) (17) can accumulate enough elastic strain energy to cause major earthquakes. Under tectonic loading, different-size earthquakes resulting from different-size cracking events of the strong segments begin to occur from their crack-initiation points, i.e., earthquakes are not limited to cracking events at their PSPs. This is consistent with the seismic phenomena that abundant earthquakes are daily observed worldwide. Herein, we refer to such a segment that dominates the generation of tectonic earthquakes as a "locked segment." As examples of intraplate seismic zones, five primary structures (Fig. 2*A*) fall under this term: a rock bridge, an asperity, a strong junction of two intersecting faults, a locked patch within a creeping segment, and a block bounded by secondary faults. In an interplate seismic zone involving a subducting plate and an overriding plate (Fig. 2*B*), because shallow-, intermediate-, and deep-focus earthquakes are mostly generated from the subducting plate (18, 19), we infer that the locked segments, including the asperities distributed along the interface between the two plates (20), are attached to the subducting plate and are distributed inside it in certain ways. These natural locked segments are significantly different from rocks studied in the laboratory in terms of geometric features and tectonic strain rates (21). For example, geological strain rates usually range from $10^{-15}$ to $10^{-14}$/s (22), which is far smaller than the common creep strain rates of $10^{-8}$ to $10^{-5}$/s measured in the laboratory (23, 24). Seismogenic locked segments are characterized by their large size and flat shape, and they are subjected to shear loading or stress corrosion at an extremely low rate, along with high temperatures and high confining pressures. These conditions make seismogenic locked segments highly heterogeneous and less brittle.

**Precursor seismicity pattern of locked segment**

Several laboratory experiments (25, 26) have demonstrated that even under high confining pressure and temperature, a stressed rock specimen can generate several AEs and fail in a less brittle manner. The premonitory AE activity of a specimen, particularly an event or energy rate at the VEP, can be enhanced if the following conditions are met: (i) the specimen is highly heterogeneous (27), (ii) it has a small height to diameter ratio (28), and (iii) it is subjected to relatively slow loading (29). This trend implies that the precursor seismicity of a large slab-shaped seismogenic locked segment may be unique under similar loading conditions that make the segment less brittle. Indeed, the mechanical behavior of a seismogenic locked segment, particularly over the long time scales relevant to large-scale tectonic deformation, cannot be modeled realistically in laboratory tests (21). Nevertheless, field-scale quasi-creep experiments on a locked-segment-like specimen may provide valuable clues in the search for precursor AE activity or seismicity patterns.

An in-situ direct shear test (30) on a large-scale slab-shaped rock block with joints and a loosening seam was performed with slow stress-stepping loading. The geometry and heterogeneity of this block, as well as the loading conditions imposed, were somewhat analogous to those of a natural locked segment. The data in Fig. 3, reprinted from Ishida, Kanagawa and Kanaori (30), show that the AE event rate rose sharply before the rupture of the block, as expected. This jump in AEs likely corresponds to the emergence of a VEP in a locked segment in the field. After the jump, AEs became relatively inactive until the locked segment was damaged to the PSP.

The earth, prone to quakes, is a unique natural laboratory for studying the quake precursor pattern and evolutionary rule. We have observed an interesting phenomenon in a well-defined seismic zone, an area with internally connected seismicity (see *Materials and Methods*), in its current seismic period, exemplified by the Tangshan seismic zone (Fig. 4*A* or No. 26 in Figs. S1 and S2). A characteristic earthquake (CE), i.e., a major earthquake, and a sequence of smaller subsequent earthquakes, which are defined as pre-shocks, always arise prior to another CE (Fig. 4*B*) (see Table 1 or below for names of such earthquakes). Note that the concept of CEs employed herein is wholly different from previous uses of the term, which typically indicates earthquakes with a similar magnitude on a given fault segment that recur at regular intervals (31). A seismicity series comprising two adjacent CEs and pre-shocks between them seemingly resembles the AE activity between the VEP and PSP shown in Fig. 3. In the following analysis, we hypothesize that CEs occur at the VEPs and PSPs of locked segments.

A single seismic period in each seismic zone can be defined by the sequential macroscopic rupture of all locked segments in ascending order of bearing capacity (low to high). As we hypothesize, when a locked segment is damaged enough that its VEP and PSP are reached, two CEs will appear at the two points, respectively. A seismic zone may have undergone many cycles of seismic periods (Fig. 5) over a long historical course, in each of which the last CE can be referred to as a mainshock. Hence, only those earthquakes occurring exclusively between the mainshock and the onset of the next seismic period are aftershocks.

In the following sections, we describe how we used a mechanical model, magnitude constraint conditions, and case studies drawn from historical data to identify pairs of CEs that occur sequentially at a VEP and a PSP of natural locked segments.

**Mechanical model, magnitude constraint conditions, and identification criterion for mainshock**

Several statistical analyses (32) have revealed that the ratios of peak strain to volume-expansion strain for rock specimens under uniaxial compression are approximately constant. This constancy probably indicates a mechanical link between the two points. To study the correlation between the shear strain at the VEP and PSP of a locked segment along a seismogenic fault or a slope slip surface, we have developed a mechanical model that couples a one-dimensional renormalization group model (16, 33, 34) with a strain-softening constitutive model (16, 17) based on the Weibull distribution (35). This mechanical model (16, 17) is expressed as

$$\frac{\varepsilon_\mathrm{f}}{\varepsilon_\mathrm{c}} = \left(\frac{2^m - 1}{m \ln 2}\right)^{\frac{1}{m}}, \qquad [1]$$

where $\varepsilon_\mathrm{c}$ and $\varepsilon_\mathrm{f}$ are the shear strain values at the VEP and the PSP of a locked segment and $m$ is the Weibull shape parameter. The shape parameter characterizes both the heterogeneity of the rock and its mechanical response to loading conditions, and hence is regarded as a common brittleness index. A smaller $m$-value indicates that the rock is less brittle. Because the $m$-values are low for a natural locked segment and lie reasonably within the range of 1.0 to 4.0 (17), and the ratio $\varepsilon_\mathrm{f} / \varepsilon_\mathrm{c}$ is insensitive to variations in the *m*-value, the ratio can be approximated as a constant 1.48, which is the average value of $\varepsilon_f / \varepsilon_c$ within $m = [1.0, 4.0]$. Thus, equation [1] is approximated as

$$\varepsilon_f = 1.48\varepsilon_c \qquad [2]$$

to predict the critical shear strain at the PSP of a locked segment. The model does not require us to determine the specific mechanical parameters of the locked segment in question because we use the constant 1.48.

As a certain seismic zone includes multiple locked segments, the relation (16) derived from equation [2] is as follows:

$$\varepsilon_f(k) = 1.48^k \varepsilon_c, \qquad [3]$$

where $\varepsilon_c$ and $\varepsilon_f(k)$ are the shear strain values corresponding to the VEP of the first locked segment and to the PSP of the *k*th locked segment, respectively. Note that the shear strain at the PSP of the *k*th locked segment is approximately equal to the shear strain at the VEP of the (*k* + 1)th locked segment, indicating that the *k*th point is both the PSP of the *k*th locked segment and the VEP of (*k* + 1)th locked segment. In other words, when the *k*th locked segment is damaged to its PSP, the load applied to it will be mostly transferred to the (*k* + 1)th locked segment, and the load transfer with a negligible shear strain increment can make the latter evolve to reach its VEP, where the CE does not occur at the PSP of the former but at the VEP of the latter (36). To test the reliability of the model, we have conducted a retrospective analysis of several landslides with locked segments, and we obtained satisfactory results (17).

Based on the assumptions that the shear strain within a locked segment is uniformly distributed and the unloading shear module during a stress drop is approximately equal to the shear elastic module of the locked segment (37), the cumulative Benioff strain (CBS) (38) can be substituted for shear strain in equation [3]. Thus, equation [3] becomes

$$S_f(k) = 1.48^k S_c, \qquad [4]$$

where $S_c$ is the CBS value measured at the VEP of the first locked segment following the error correction (see below), and $S_f(k)$ is the CBS value predicted at the PSP of the $k$th locked segment.

Because a certain seismic zone ordinarily contains seismic events irrelevant to the brittle failures of locked segments, the minimum validity magnitude $M_v$ (39) is set such that an analysis determines the threshold magnitude of the locked segments' cracking and faulting events based on the earthquake catalogue of that zone. $M_v$ is usually not less than the minimum completeness magnitude $M_c$ (40). Particularly before the 1900s, earthquakes whose magnitudes are not smaller than $M_v$ preceding the VEP of the first locked segment were usually not completely recorded. Hence, an initial error in the CBS may be introduced. Therefore, we propose an error-correction expression using equation [4]:

$$\Delta = \frac{S_f^*(1) - 1.48 S_c^*}{0.48}, \qquad [5]$$

where $\Delta$ denotes the error, and $S_c^*$ and $S_f^*(1)$ are the uncorrected recorded CBS values at the VEP and PSP of the first locked segment in the zone, respectively.

If there are $n$ CEs in a seismic zone and its current seismic period, then the $n$th CE (the last CE) is a mainshock. According to the energy conversion and allocation principle during the damage process of a locked segment (37) and comprehensive statistical analysis (41-43) of all the earthquake cases conducted worldwide, we found that the lower-limit constraints for the magnitudes of two adjacent CEs or several successive CEs from the $i$th to the $(i + j)$th satisfy

$$M_i - M_{i+j} \leq 0.2 \ (M_{i+j} < M_i \ (1 \leq i \leq n - 2, j \geq 1, \text{ and } i + j \leq n - 1)), \qquad [6]$$

the upper-limit constraints for the magnitudes of two adjacent CEs satisfy

$$M_{i+1} - M_i \leq 0.5 \ (M_{i+1} \geq M_i \ (1 \leq i \leq n - 2)), \qquad [7]$$

and the upper-limit constraints for the magnitude of pre-shocks between two adjacent CEs satisfy

$$M_P \leq \min(M_{i+1}, M_i) - 0.2 \ (1 \leq i \leq n - 1), \qquad [8]$$

where $M_i$, $M_{i+j}$, and $M_{i+1}$ denote the magnitudes of the $i$th CE, $(i+j)$th CE, and $(i+1)$th CE, respectively, and $M_P$ denotes the magnitude of pre-shocks. The trend for the magnitudes of several successive CEs to satisfy formula [7] is relatively common, as observed in the Tangshan seismic zone. Double earthquakes with equal or similar magnitudes within a short time are viewed as an equivalent earthquake whose seismic energy equals the sum of their released energy.

Once the magnitude scale has been standardized (see *Materials and Methods*), formulas [6] and [7] can be used to revise and predict the magnitudes of CEs (except for a mainshock), whereas formula [8] can be used to revise and predict the magnitudes of large pre-shocks. Herein, these formulas are referred to as the magnitude constraint conditions.

Considering several inaccurate or controversial parameters in the earthquake catalogue of a certain seismic zone, particularly magnitude parameters, we revised the magnitudes (41-43) using the constraint conditions on the basis of several catalogues compiled by individuals and international agencies and research results. The revision rules considerably reduce the uncertainty of our earthquake case studies.

Once the last locked segment with the highest bearing capacity is damaged to its PSP, a mainshock will occur within the same seismic zone and period. We can identify the mainshock by the following relation (37):

$$M_n - M_{n-1} > 0.5, \quad [9]$$

where $M_n$ and $M_{n-1}$ denote the magnitudes of the mainshock and the CE at the VEP of the last locked segment, respectively.

Given the error in the magnitude measurement, when the magnitude difference between two adjacent CEs is around 0.5, judging whether the $(n+1)$th CE is a mainshock based on formula [9] is difficult. Therefore, we recommend further investigations into the seismicity characteristics in a certain seismic zone following the CE. If the subsequent earthquakes whose magnitudes are not less than $M_v$ occur randomly and frequently over a long period, the CE is not a mainshock and its magnitude must be revised using the aforementioned revision rules.

**Case studies**

Earthquakes have been recorded across a nearly 3800-year-long period in the intraplate Tangshan seismic zone, which is clearly bounded by several significant deep-seated regional faults (44) (Fig. 4*A*). Therefore, the Tangshan seismic zone is a strong candidate for ascertaining seismicity characteristics over time and validating our mechanical model. The earthquake catalogues in the Tangshan seismic zone before and after 1900 are drawn from Song, Zhang and Liu (45) (see *Materials and Methods*) and the China Earthquake Data Center (CEDC). We revised the magnitudes of several large historical events recorded before the invention of instrumentation (41) using the aforementioned revision rules. Twelve $M_S \geq 7.0$ earthquakes (Table 1) have been recorded in this zone, five of which qualify as CEs under the definition proposed above (Fig. 4*B*): 1597-10-06 $M_S$7.5 Bohai Sea earthquake in China (CE1), 1668-07-25 $M_S$8.0 Tancheng earthquake (CE2), 1679-09-02 $M_S$7.8 Sanhe-Pinggu earthquake (CE3), 1888-06-13 $M_S$7.8 Bohai Bay earthquake (CE4) and 1976-07-27 $M_S$7.8 Tangshan earthquake (CE5). The Tangshan earthquake was the largest event recorded in the zone since 1900, and it caused heavy casualties and property damage because of the failure in predicting/forecasting it. If these earthquakes are CEs that indicate VEPs and PSPs of locked segments, their evolutions should match equation [4].

The $M_v$ in the Tangshan seismic zone is $M_S$5.0. After the various magnitudes in the record were transformed into a uniform scale (see *Materials and Methods*), the CBS of $M_S \geq 5.0$ earthquakes in this zone was calculated using a widely known formula (38). The initial error of CBS was determined using equation [5]. Then, the correlation of critical CBS values among CEs in the seismic zone can be quantified using equation [4]. The data in Fig. 4*C* and Table 2 show that the evolutions of these CEs match equation [4] very well, confirming our hypothesis. This shows that the 1597-10-06 $M_S$7.5 Bohai Sea earthquake and the 1668-07-25 $M_S$8.0 Tancheng earthquake form a pair of CEs at the VEP and PSP of the first locked segment. By parity of reasoning, the Tancheng earthquake and the 1679-09-02 $M_S$7.8 Sanhe-Pinggu earthquake form a pair of CEs at the second locked segment in the zone. The first earthquake of each pair of CEs is an identified precursor for the second earthquake in the pair. Following the 1969-07-18 $M_S$7.4 Bohai Sea earthquake (pre-shock), the PSP of the fourth locked segment was almost reached; hence, the fifth CE corresponding to the PSP of the segment, the $M_S$7.8 Tangshan earthquake, occurred about seven years later, indicating that it is a predictable CE. According to formula [9], this CE was not a mainshock; therefore, a further CE is expected.

Additionally, we tested our hypothesis with a retrospective analysis of the interplate Hokkaido seismic zone (Fig.6*A* or No. 38 in Fig. S1), which lies in the Okhotsk plate and is constrained by plate boundaries. Shallow-, intermediate- and deep-focus earthquakes have been recorded in this zone. The earthquake catalogues in this zone before and after 1900 were obtained from Song, Zhang and Liu (45) and the USGS National Earthquake Information Center (NEIC). Three recognizable megathrust CEs are found: the 1898-06-05 $M_{uk}$8.7 Japan Trench earthquake (CE1), 1952-11-04 $M_W$8.9 earthquake (CE2) off the eastern coast of Kamchatka and 2011-03-11 $M_W$9.0 Tohoku-Oki, Japan earthquake (CE3). Using the same method as we did for the above analysis, we find that pairs of these CEs (Fig. 6*B* and Table 2) also satisfy equation [4] without modifications of the catalogue. This agreement indicates that brittle failures of locked segments led to the shallow-, intermediate- and deep-focus earthquakes whose magnitudes are not smaller than $M_v$ ($M_v = M_W7.0$) recorded in this zone.

Intermediate- and deep-focus earthquakes, which are highly sensitive to the temperature of a subducting plate (46), are most often observed from mature subduction zones in cold subducting plates (47). These earthquakes have occurred in the coldest central portions of subduction zones where the temperature may not exceed 600°C (48, 49). Such a condition enables the locked segments within the portions to produce brittle failures.

Equation [4] has been corroborated by our retrospective analysis of earthquake data (e.g., Figs. S3 to S6) in 62 seismic zones (Figs. S1 and S2) covering the circum-Pacific seismic belt and the Eurasia seismic belt. The data were drawn from the earthquake catalogue published (see *Materials and Methods*) and revised for consistency. The pattern we have identified applies regardless of focal depth, indicating that the precursor pattern is universal. Consequently, we conclude that a CE at the VEP of locked segment reliably precedes the rupture of the segment that leads to a CE at the segment's PSP.

**Discussion and conclusions**

As the damage evolution from the VEP of a locked segment to its PSP usually takes decades or even hundreds of years, the earthquake precursor we have identified can be considered as a long-term precursor. The precursor seismicity pattern and our quantitative mechanical model permit the prediction of CEs within a given seismic zone, except for the first two. Using equation [4] and formulas [6] to [9], one can identify CEs and predict the critical CBS value and magnitude range of a future CE together with the upper-limit magnitude of pre-shocks before the said CE within a well-defined seismic zone.

The reason that most researchers have failed to find such a precursor pattern through conventional rock rupture tests is that intact small-scale cylindrical specimens, with a height to diameter ratio of 2:1 under rapid loading, were characterized by low heterogeneity and high brittleness levels that are considerably different from those of natural locked segments. Therefore, we suggest that future testing and numerical modeling to obtain detailed insights into the damage mechanism of locked segments and rocks should focus on these properties.

Recently, we developed methods for identifying the first foreshock, i.e., a special pre-shock that serves as an indicator of approaching or reaching the critical state of a CE, within a defined seismic zone (37). For example, the 1969-07-18 $M_S$7.4 Bohai Sea earthquake mentioned above was the first foreshock preceding the predictable CE, the $M_S$7.8 Tangshan earthquake. When the recorded CBS value of a seismic zone approaches or reaches the predicted critical CBS value, one can estimate the time window of a future CE according to the lag time between the previous foreshocks and CEs within the zone. Our analysis of all the earthquake cases relating to intraplate seismic zones shows the existence of a relatively long quiet period of seismicity between the first foreshock and subsequent CE, suggesting that the physical state of the corresponding locked segment remains almost unchanged during this period; therefore, the identifiable physical precursors, such as anomalies in wave velocity, and electrical and magnetic signals, are absent.

The case studies reveal that the earthquake generation mechanism is attributed to the brittle failure of stressed locked segments, which macroscopically rupture one-by-one in the order of bearing capacity from low to high, rather than elastic rebound (50) or stick–slip (51), as has been widely assumed. Different-size cracking events generated from the stressed locked segments correspond to quite a few pre-shocks and rare CEs. Among these, the former are unpredictable random events based on current knowledge, regardless of size; however, evolutions of the latter have definite physical meanings and follow a deterministic rule, thereby ensuring their predictability.

When a locked segment is broken, the load applied to it will be mostly transferred to the next locked segment within the same seismic zone, thereby accelerating its failure process. As the locked segments fail one-by-one, the remaining segments will bear more and more transferred load; thus, an increasingly accelerating seismicity trend within this zone is inevitable, as illustrated in Figs. 4*C* and 6*B* and Figs. S3 to S6. Once the last locked segment with the highest bearing capacity is damaged to its PSP, a mainshock will occur within the same seismic zone and period. This demonstrates that the earthquakes preceding the rupture of the last locked segment are neither mainshocks nor aftershocks, as argued by previous researchers. As each seismic zone evolves towards the critical state of the mainshock, the larger pre-shocks and CEs generated from the stressed locked segments with higher bearing capacity will appear more frequently. Thus, we recommend that countries and regions in seismically active areas should improve the ability to prevent and mitigate earthquake disasters as early as possible.

Our work has revolutionized the understanding of the earthquake generation process and basic concepts, such as mainshock, aftershock, seismic zone, and seismic period. These concepts stem from our understanding of earthquake mechanisms and their rationality has been validated in our case studies. Owing to the special properties of locked segments, i.e., high heterogeneity and low brittleness, a universal seismicity precursor pattern and a specific rule describing the evolution of CEs for each well-defined seismic zone exist; these lay the physical foundation for earthquake predictability.

**Materials and Methods**

**Seismic zoning.** The lithosphere of the Earth can be thought of as a hierarchy of blocks with relative movement separated by faults or fault networks at different scales (52), including tectonic plates (53-57), fault blocks (58, 59), and active tectonic blocks (60, 61). Because the tectonic deformation of major faults, including basement faults, crustal faults, lithospheric faults, and translithospheric faults, is much greater than that inside a block divided by these faults, the block essentially moves as a unit, suggesting that the major faults can serve as the boundaries of the block. In this context, earthquakes within an intraplate tectonic block bounded by major faults, or by major faults and plate boundaries, are intrinsically linked, as are earthquakes within a subducting plate (an interplate tectonic block) constrained by major faults or plate boundaries, or by both. Conversely, adjacent blocks only affect the loading or unloading mode of a particular block via shearing or extruding action; they do not affect the intrinsic evolutionary rule reflected in the internal seismicity of the block in question (62). Thus, a seismic zone can be defined as an area representing the seismicity of a corresponding tectonic block. In previously published works (41-43), we confirmed that the regular evolution of seismicity with a magnitude equal to or greater than $M_v$ within a seismic zone can represent the unity of a block.

Definitely, our concept of seismic zoning highlights the mechanical interaction between fault networks within a specified seismic zone, which is different from the conventional understanding that earthquakes are related to an individual fault or a fault zone. Relying on data for the faults and plate boundaries illustrated in a seismo-tectonic map (44, 63-66), we have defined 62 seismic zones (Fig. S1) that cover the circum-Pacific seismic belt and the Eurasia seismic belt, which includes 33 seismic zones in China and its neighboring areas (Fig. S2).

**Earthquake catalogues.** For our case studies on 62 seismic zones around the world, we adopted the pre-1900 earthquake catalogue compiled by Song, Zhang and Liu (45), which is a revised collection of various global catalogues available from authority websites, published books, and documents, including $M \geq 5.0$ and $M \geq 6.0$ earthquakes from 9999 BC to 1963 AD and 1964 AD to May 2011, respectively. The post-1900 earthquake catalogue of earthquakes in China and its neighboring regions was taken mostly from the CEDC (http://data.earthquake.cn), and the post-1900 earthquake catalogue for other regions was compiled from the NEIC database (http://earthquake.usgs.gov).

**Conversion between different magnitude scales.** Magnitudes are given in the catalogues using several scales, e.g., surface wave magnitude ($M_S$), short- and long-period body wave magnitudes ($m_b$ and $m_B$), local magnitude ($M_L$), moment magnitude ($M_W$), magnitude defined by the Japan Meteorological Agency ($M_j$), magnitude calculated statistically ($M_{uk}$), and magnitudes estimated by the intensity of hazards ($M_K$) and fault area ($M_{fa}$). Before calculating CBS, all these magnitudes are transformed into a local magnitude scale, $M_L$, using the appropriate formulas (67). Subsequently, we calculate the seismic energy of each event whose magnitude is not less than $M_v$, the Benioff strain, and the CBS using well-known formulas (67).

**Acknowledgments**

We thank S. Wang and J. Yang of the Institute of Geology and Geophysics, Chinese Academy of Sciences, for their valuable comments and suggestions. This work was supported by the National Natural Science Foundation of China to S. Q. (grant 41572311 and U1704243) and L. X. (grant 41302233), and by the China Postdoctoral Science Foundation (grant 2018M640181) to H. C.

**Figures and Tables**

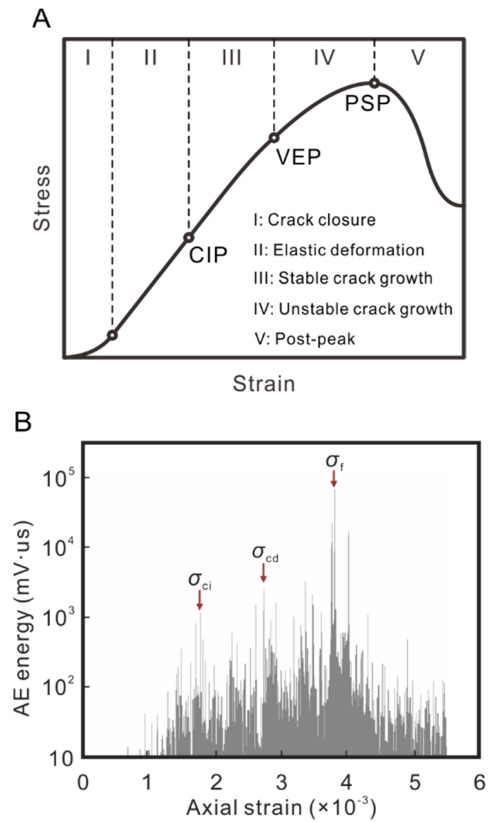

**Figure 1.** Deformation and failure process of rock under compression or shear. (*A*) Ordinary stress–strain relation of rock that comprises five phases. The phases II to V are divided by crack-initiation point (CIP), volume-expansion point (VEP), and peak-stress point (PSP). (*B*) AE energy vs. strain measured from a granodiorite specimen subjected to uniaxial compression (10). $\sigma_{ci}$, $\sigma_{cd}$, and $\sigma_f$ indicate the stress levels corresponding to CIP, VEP, and PSP of the specimen.

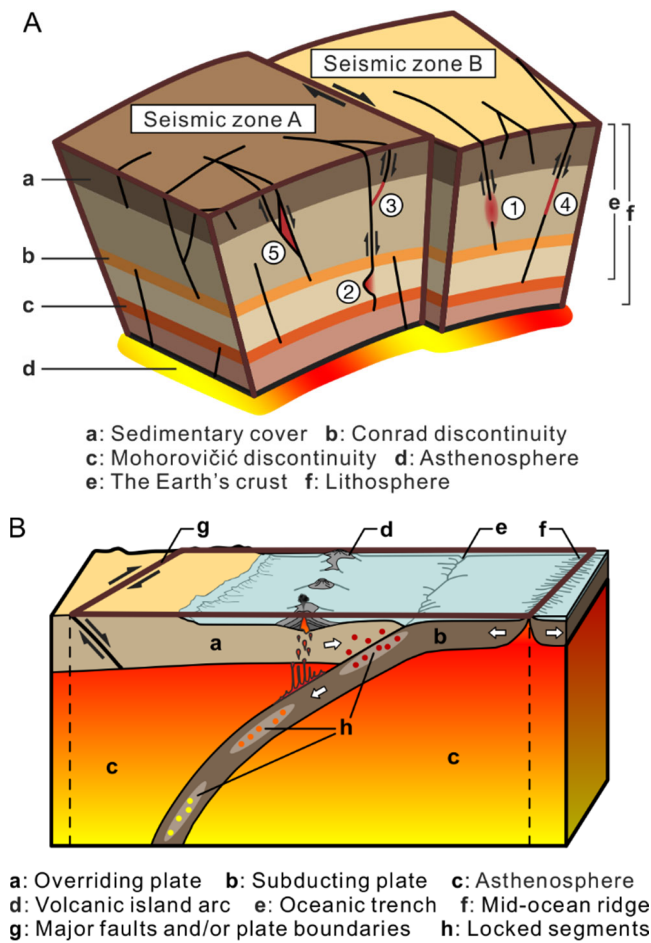

a: Sedimentary cover  b: Conrad discontinuity
c: Mohorovičić discontinuity  d: Asthenosphere
e: The Earth's crust  f: Lithosphere

a: Overriding plate  b: Subducting plate  c: Asthenosphere
d: Volcanic island arc  e: Oceanic trench  f: Mid-ocean ridge
g: Major faults and/or plate boundaries  h: Locked segments

**Figure 2.** Illustration of the concepts of seismic zones and locked segments. (*A*) Adjacent tectonic blocks whose corresponding areas are intraplate seismic zones, A (light brown) and B (yellowish), are bounded by major faults (thick auburn lines). Whether a block readily slips along a bottom discontinuity depends on the development level and friction resistance of the discontinuity and tectonic load imposed. If a block slips along a certain discontinuity, the block above the discontinuity is the specified one. Thin black lines represent the secondary faults in a seismic zone. Black arrows denote the movement direction of active faults. Red areas denoted by numerical labels refer to five primary categories of seismogenic locked segments: ① a rock bridge, ② an asperity, ③ a strong junction of two intersecting faults, ④ a locked patch in a creeping segment, and ⑤ a block bounded by secondary faults. (*B*) An interplate seismic zone (encircled by a thick closed auburn curve) is constrained by major faults or plate boundaries, or by both. Because such a zone contains the seismic events generated from a subducting plate and an overriding plate, the events from the latter, which are irrelevant to failures of locked segments, need to be removed via data processing according to the seismic zone definition (see *Materials and Methods*),

which can be achieved by introducing $M_v$. Red, orange, and yellow solid circles show shallow-, intermediate-, and deep-focus earthquakes, respectively, whereas white and black arrows denote the movement direction of the plate and fault, respectively.

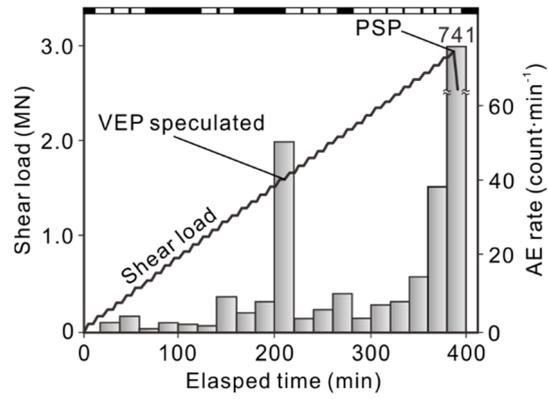

**Figure 3.** AE rate (gray bars) and shear load (zigzag line) during an in-situ direct test run (30). The last broken bar denotes an extremely high AE rate, of 741 count·min$^{-1}$, at the PSP. White and black rectangles at the top of the figure represent the periods with and without AE measurements, respectively.

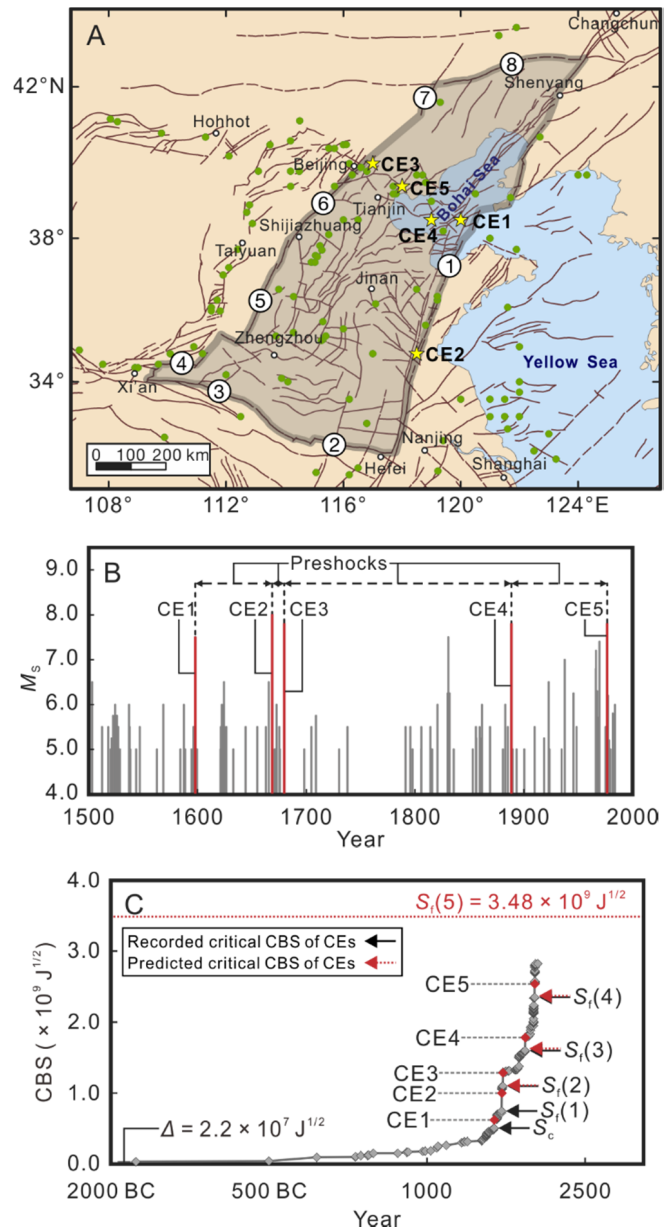

**Figure 4.** Seismicity in the Tangshan seismic zone. (*A*) Seismo-tectonic context of the seismic zone. The seismic zone (shadow area encircled by a closed gray thick curve) is bounded by several major faults (auburn curves denoted by numerical labels): ① Tanlu fault; ② Yangce-Gushi-Feizhong fault; ③ Tieluzi-Luanchuan-Nanzhao fault; ④ Huashan Mountain piedmont fault; ⑤ Taihang Mountain large fault; ⑥ Taihang Mountain piedmont fault; ⑦ Balihan fault; ⑧ Chifeng-Kaiyuan fault. Green solid circles show the locations of $M_S \geq 6.0$ earthquakes, of which five CEs (CE1 to CE5) are marked with yellow stars. (*B*) Time series of $M_S \geq 5.0$ earthquakes. To clarify the CEs, the seismic data between 1500 and 2000 are selected (no $M_S > 7.0$ earthquakes were recorded before 1500 and after 2000). Red bars

denote CE1 to CE5. (*C*) Time variations of CBS of $M_S \geq 5.0$ earthquakes from 1767 BC to 19 November 2015, in the seismic zone and its current seismic period after introducing the initial error Δ. Red diamonds denote the CEs, and gray diamonds denote the other $M_S \geq 5.0$ earthquakes. Table 2 lists the recorded and predicted critical CBS values. The predicted critical CBS value for an expected future CE, calculated from equation [4], is marked with a dotted red line.

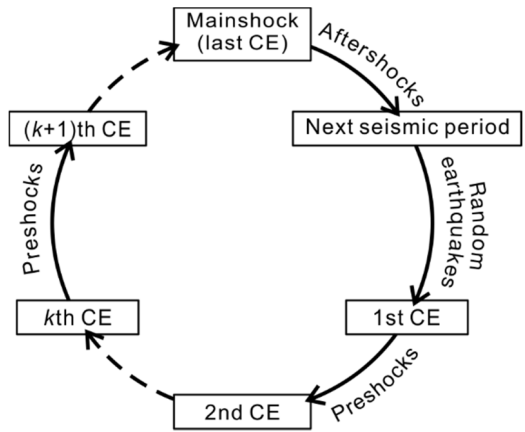

**Figure 5.** Cycle of seismic period for a defined seismic zone.

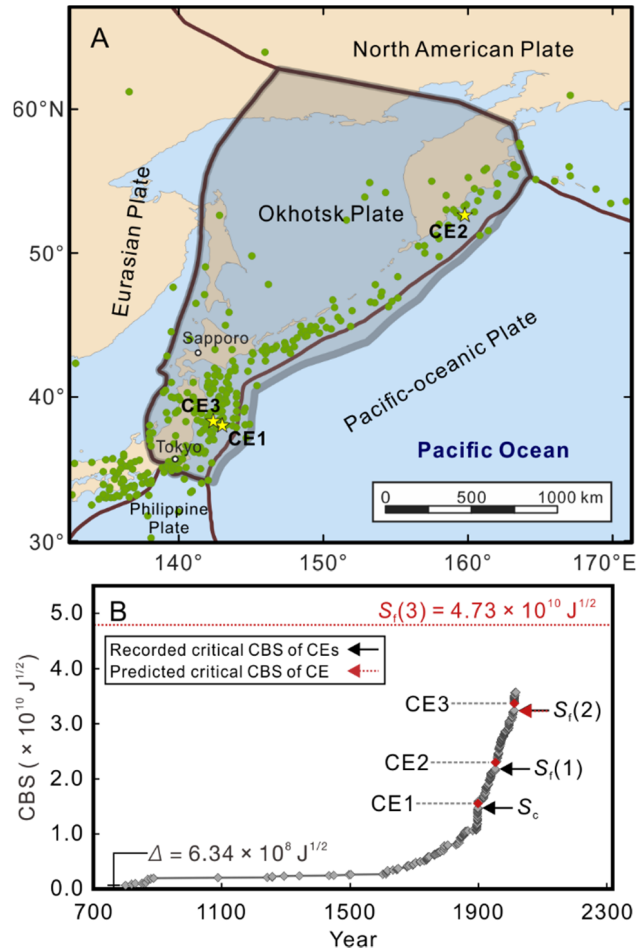

**Figure 6.** Seismicity in the Hokkaido seismic zone. (*A*) Seismo-tectonic context of the seismic zone. The seismic zone (shadow area encircled by a closed gray thick curve) is attached to the Okhotsk Plate and is constrained by the plate boundaries (auburn curves) of four neighboring plates. Green solid circles plot the locations of $M_W \geq 7.0$ earthquakes, of which three CEs (CE1 to CE3) are marked with yellow stars. (*B*) Time variations of the CBS of $M_W \geq 7.0$ (i.e., $M_v = M_W 7.0$) earthquakes from 15 February 144, to 2 February 2016, in the Hokkaido seismic zone and its current seismic period after introducing the initial error $\Delta$. Red diamonds mark the CEs, and gray diamonds mark the other $M_W \geq 7.0$ earthquakes. Table 2 lists the recorded and predicted critical CBS values. The predicted critical CBS value for an expected future CE, calculated from equation [4], is marked with a dotted red line.

**Table 1.** $M_S \geq 7.0$ earthquakes in the Tangshan seismic zone of which five CEs are highlighted by gray background.

| No. | Date | Latitude and Longitude (°) | Magnitude ($M_S$) | Location |
|---|---|---|---|---|
| 1 | 70 BC-06-01 | 36.30 N, 119.20 E | 7.0 | Zhucheng, Shandong |
| 2 | 1597-10-06 | 38.50 N, 120.00 E | 7.5 | Bohai Sea |
| 3 | 1668-07-25 | 34.80 N, 118.50 E | 8.0 | Tancheng, Shandong |
| 4 | 1668-07-26 | 36.40 N, 119.20 E | 7.0 | Anqiu, Shandong |
| 5 | 1679-09-02 | 40.00 N, 117.00 E | 7.8 | Sanhe-Pinggu, Hebei |
| 6 | 1830-06-12 | 36.40 N, 114.30 E | 7.5 | Cixian, Hebei |
| 7 | 1888-06-13 | 38.50 N, 119.00 E | 7.8 | Bohai Bay |
| 8 | 1937-08-01 | 35.20 N, 115.30 E | 7.0 | Heze, Shandong |
| 9 | 1966-03-22 | 37.50 N, 115.10 E | 7.2 | Xingtai, Hebei |
| 10 | 1969-07-18 | 38.20 N, 119.40 E | 7.4 | Bohai Sea |
| 11 | 1976-07-27 | 39.40 N, 118.00 E | 7.8 | Tangshan, Hebei |
| 12 | 1976-07-28 | 39.70 N, 118.50 E | 7.5 | Luanxian, Hebei |

**Table 2.** Recorded and predicted critical CBS values corresponding to the predictable CEs in Tangshan (Fig. 4*C*) and Hokkaido (Fig. 6*B*) seismic zones.

| Names of seismic zones | CE | Recorded CBS value ($J^{1/2}$) | Predicted CBS value ($J^{1/2}$) |
|---|---|---|---|
| Tangshan | CE3 | $1.10 \times 10^9$ | $1.11 \times 10^9$ |
|  | CE4 | $1.60 \times 10^9$ | $1.63 \times 10^9$ |
|  | CE5 | $2.35 \times 10^9$ | $2.37 \times 10^9$ |
| Hokkaido | CE3 | $3.24 \times 10^{10}$ | $3.23 \times 10^{10}$ |

Supplementary Information for

Universal precursor seismicity pattern before locked-segment rupture

Hongran Chen[a, b], Siqing Qin[a, b, c], Lei Xue[a, b], Baicun Yang[a, b, c] and Ke Zhang [a, b, c]
[a]Key Laboratory of Shale Gas and Geoengineering, Institute of Geology and Geophysics, Chinese Academy of Sciences, Beijing, China. [b]Innovation Academy for Earth Science, Chinese Academy of Sciences, Beijing, China. [c]College of Earth and Planetary Sciences, University of Chinese Academy of Sciences, Beijing, China.

*Siqing Qin
Email: qsqhope@mail.iggcas.ac.cn

**This PDF file includes:**

    Figures S1 to S6

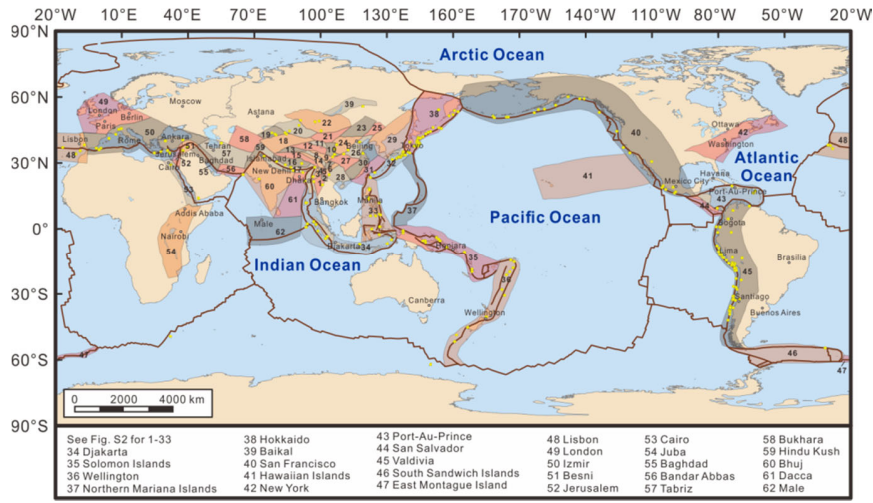

**Fig. S1.** Seismic zoning map including 62 zones (color shadow areas defined by semi-transparent gray thin curves) worldwide (adapted from Qin et al. (1)). Numerical labels in black bold font are attached to each zone; the zone names corresponding to these numbers are listed in the lower table. Data concerning the tectonic plate boundaries (auburn curves) were taken from the NEIC database (http://earthquake.usgs.gov, last accessed on September 11, 2014). Yellow solid circles show the locations of $M \geq 8.0$ earthquakes.

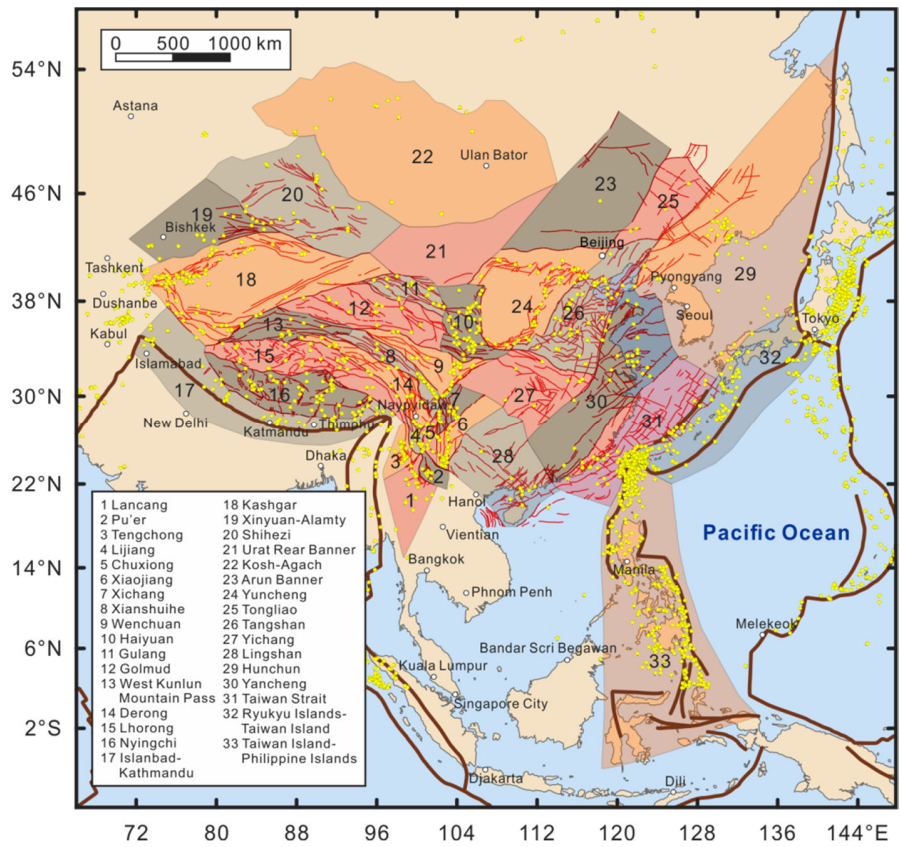

**Fig. 2.** Seismic zoning map including 33 zones (color shadow areas defined by semi-transparent gray thin curves) in China and its neighboring areas (adapted from Qin et al. (2)). Each zone is labeled with a number in black font; the zone names corresponding to these numbers are listed in the lower-left table. The fault data (red curves) are from the Map of Active Tectonics in China (3) and the data of tectonic plate boundaries (thick auburn curve) are from the NEIC (http://earthquake.usgs.gov, last accessed on September 11, 2014). Yellow solid circles show the locations of $M \geq 6.0$ earthquakes.

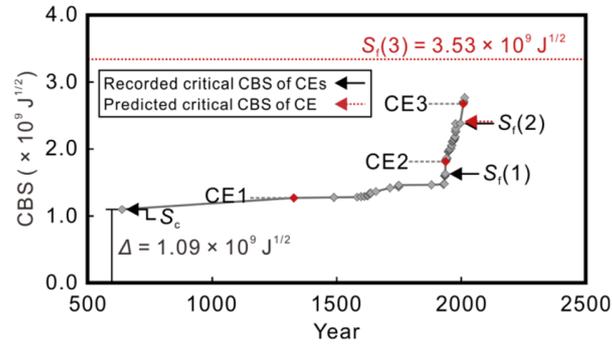

**Fig. S3.** Time variations of CBS of $M_S \geq 5.5$ (i.e. $M_v = M_S 5.5$) earthquakes from February 14, 638 to August 21, 2017 in the Wenchuan seismic zone (No. 9 in Figure S1 and S2) and its current seismic period after introducing the initial error $\Delta$. Red diamonds denote the CEs and gray diamonds denote the other $M_S \geq 5.5$ earthquakes. The CEs in this zone are the 1327-09 $M_S 7.75$ Tianquan earthquake (CE1) in Sichuan, 1937-01-07 $M_S 7.8$ Maduo earthquake (CE2) in Qinhai, and 2008-05-12 $M_S 8.1$ Wenchuan earthquake (CE3) in Sichuan. The predicted critical CBS value for an expected future CE, calculated from equation [4], is marked with a dotted red line.

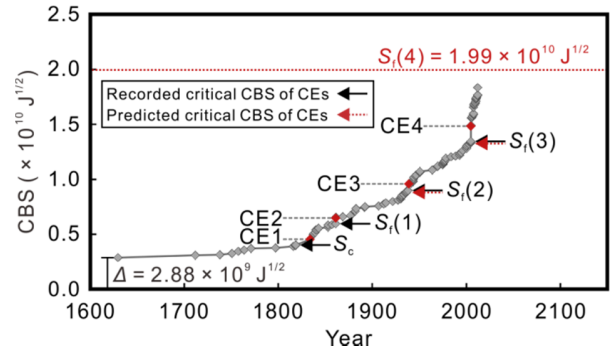

**Fig. S4.** Time variations of CBS of $M_W \geq 7.0$ (i.e. $M_v = M_W 7.0$) earthquakes from August 1, 1629 to February 27, 2015 in the Djakarta seismic zone (No. 34 in Figure S1) and its current seismic period after introducing the initial error $\Delta$. Red diamonds denote CEs, and gray diamonds denote the other $M_W \geq 7.0$ earthquakes. The CEs in this zone are the 1818-11-08 $M_S 8.5$ earthquake (CE1) in Bali Sea, 1861-02-16 $M_S 8.5$ Simuk, Indonesia earthquake (CE2), 1938-02-01 $M_W 8.5$ Banda Sea earthquake (CE3), and 2004-12-26 $M_W 9.0$ Simeulue Island earthquake (CE4). The predicted critical CBS value for an expected future CE, calculated from equation [4], is marked with a dotted red line.

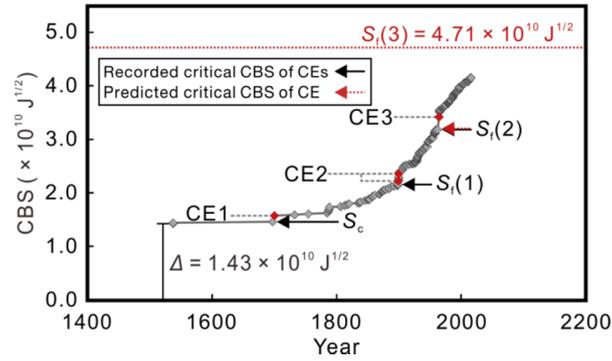

**Fig. S5.** Time variations of CBS of $M_W \geq 7.0$ (i.e. $M_v = M_W 7.0$) earthquakes from December 20, 1523 to February 24, 2016 in the San Francisco seismic zone (No.40 in Figure S1) and its current seismic period after introducing the initial error $\Delta$. Red diamonds denote the CEs and gray diamonds denote the other $M_W \geq 7.0$ earthquakes. The CEs in this zone are the 1700-01-26 $M_S 9.0$ earthquake (CE1) near coast of west Oregon, USA, 1899-01-04 $M_{uk} 8.6$ and 1899-09-10 $M_W 8.6$ double earthquakes occurring in Guerrero-Oaxaca, Mexico, and Gulf of Alaska, USA, respectively, whose releasing energy was equivalent to that of a $M_W 8.8$ event (CE2), and 1964-03-28 $M_W 9.3$ Alaska earthquake (CE3). The predicted critical CBS value for an expected future CE, calculated from equation [4], is marked with a dotted red line.

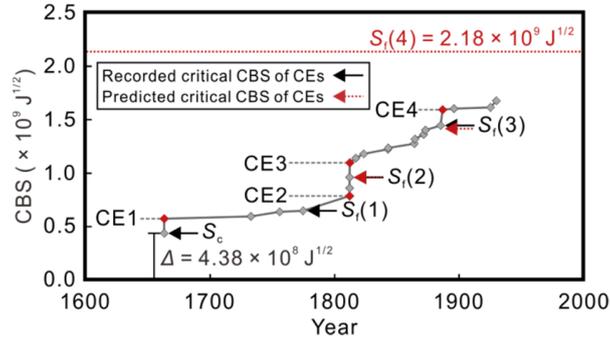

**Fig. S6.** Time variations of CBS of $M_W \geq 6.0$ (i.e. $M_v = M_W 6.0$) earthquakes from December 27, 1568 to February 15, 2014 in the New York seismic zone (No.42 in Figure S1) and its current seismic period after introducing the initial error $\Delta$. Red diamonds denote the CEs and gray diamonds denote the other $M_W \geq 6.0$ earthquakes. The CEs in this zone are the 1663-02-05 $M_S 7.6$ St Lawrence Valley, Canada, earthquake (CE1), 1811-12-16 $M_S 7.6$ earthquake (CE2) and 1812-02-07 $M_S 7.6$ earthquake (CE3) in New Madrid, USA, and 1886-09-01 $M_W 7.7$ Charleston, USA, earthquake (CE4). The predicted critical CBS value for an expected future CE, calculated from equation [4], is marked with a dotted red line.